\documentclass{rmaa}
\usepackage{multirow}

\newcommand{\lloii}{$\lambda\lambda$~7320, 7330}
\newcommand{\llsii}{$\lambda\lambda$~4068, 4076}
\newcommand{\sii}{[\ion{S}{2}]}
\newcommand{\oii}{[\ion{O}{2}]}
\newcommand{\siii}{[\ion{S}{3}]}
\newcommand{\oiii}{[\ion{O}{3}]}
\newcommand{\toii}{\ensuremath{T_\mathrm{e}(\ion{O}{2})}}
\newcommand{\tnii}{\ensuremath{T_\mathrm{e}(\ion{N}{2})}}
\newcommand{\tsii}{\ensuremath{T_\mathrm{e}(\ion{S}{2})}}
\newcommand{\toiii}{\ensuremath{T_\mathrm{e}(\ion{O}{3})}}
\newcommand{\tsiii}{\ensuremath{T_\mathrm{e}(\ion{S}{3})}}
\newcommand{\beq}{\begin{equation}}
\newcommand{\eeq}{\nonumber\end{equation}}
\newcommand{\beqar}{\begin{eqnarray}}
\newcommand{\eeqar}{\nonumber\end{eqnarray}}
\newcommand{\eden}{\ensuremath{N_\mathrm{e}}}
\newcommand{\etem}{\ensuremath{T_\mathrm{e}}}

\newcommand{\neb}{\ensuremath{\oiii\,\lambda\,5007}}
\newcommand{\odos}{\ensuremath{\oii\,\,\lambda\,3727}}

\newcommand{\noIInh}{\ensuremath{N(\mathrm{O}^{+})/N(\mathrm{H}^+)}}
\newcommand{\noIIInh}{\ensuremath{N(\mathrm{O}^{++})/N(\mathrm{H}^+)}}

\title{Chemical Abundances of NGC~5461 and NGC~5471 Derived from Echelle
 Spectrophotometry}

\author{V.~Luridiana,\altaffilmark{1}
        C.~Esteban,\altaffilmark{2}
        M.~Peimbert,\altaffilmark{3}
        and A.~Peimbert\altaffilmark{3} \medskip}
 
\altaffiltext{1}{Instituto de Astrof\'\i sica de Andaluc\'\i a (CSIC),
              Spain.}
\altaffiltext{2}{Instituto de Astrof\'\i sica de Canarias, Spain.}
\altaffiltext{3}{Instituto de Astronom\'\i a, Uni\-ver\-sidad
Nacio\-nal Aut\'o\-noma de M\'exico, M\'exico.}
   
\shortauthor{Luridiana et al.}
\shorttitle{Chemical abundances of NGC~5461 and NGC~5471} 

\fulladdresses 
{\item C\'esar Esteban: Instituto de Astrof\'\i sica de Canarias,
v\'{\i}a L\'actea s/n E-38200, La
Laguna, Tenerife, Spain (cel@ll.iac.es).
\item Valentina Luridiana: Instituto de Astrof\'{\i}sica de Andaluc\'{\i}a,
(CSIC), Apdo. de Correos 3004, (Camino Bajo de Huetor 24), Granada,
E-18080, Spain (vale@iaa.es).
\item Manuel Peimbert and Antonio Peimbert: Instituto de Astronom\'\i a, UNAM,
Apartado Postal 70-264, 04510 M\'exico D.F., M\'exico
 (peimbert, antonio@astroscu.unam.mx).}

\listofauthors{V.~Luridiana, C.~Esteban, M.~Peimbert, \& A.~Peimbert}
\indexauthor{Luridiana,~V.}
\indexauthor{Esteban, C.}
\indexauthor{Peimbert, M.}
\indexauthor{Peimbert, A.}

\SetVolume{38} \SetFirstPage{97} \SetYear{2002}
\ReceivedDate{2002 February 15}
\AcceptedDate{2002 April 16}

\addkeyword{H~II regions}
\addkeyword{ISM: abundances}
\addkeyword{ISM: individual (NGC~5461, NGC~5471)}

\resumen{ Presentamos espectrofotometr\'\i a de alta resoluci\'on de
  NGC~5461 y NGC~5471, dos regiones \ion{H}{2} gigantes en la galaxia
  M101. Los datos se obtuvieron con el teles\-copio de 2.1~m del
  Observatorio Astron\'omico Nacional en San Pedro M\'artir, Baja
  California. Medimos las intensidades de un conjunto de l\'\i neas de
  recombinaci\'on de hidr\'ogeno y de helio, as\'{\i} como l\'\i neas
  prohibidas de un gran n\'umero de iones. Calculamos las condiciones
  f\'\i sicas en las dos nebulosas y calculamos las abundancias qu\'\i
  micas totales tomando en cuenta las abundancias i\'onicas observadas
  as\'{\i} como las no observadas; estas \'ultimas las estimamos a
  partir de factores de correcci\'on de ionizaci\'on
  ($\mathit{icf}$'s). Para NGC~5461 los $\mathit{icf}$'s se basan en
  un modelo detallado de fotoionizaci\'on dise\~nado espec\'\i
  ficamente para este objeto (Luridiana \& Peimbert 2001), mientras
  que para NGC~5471 se obtuvieron a partir de un modelo de
  fotoionizaci\'on para NGC~2363 (Luridiana, Peimbert, \& Leitherer
  1999), regi\'on que muestra un grado de ionizaci\'on muy semejante
  al de NGC~5471.  Los $\mathit{icf} s$ as\'{\i} determinados los
  comparamos con aquellos que se obtienen a partir de las f\'ormulas
  de Mathis \& Rosa\@ (1991). Dicha comparaci\'on muestra importantes
  discrepancias para algunos de los elementos incluyendo nitr\'ogeno,
  ne\'on y cloro.  }

\abstract{We present high resolution spectroscopic data of the two
  giant extragalactic \ion{H}{2} regions NGC~5461 and NGC~5471 in
  M101, which have been obtained with the 2.1-m telescope of the
  Observatorio Astron\'omico Nacional in San Pedro M\'artir, Baja
  California.  We measured the intensities of several H and He
  {recombination} lines, and of forbidden lines of a large number of
  ions.  We calculate the physical conditions in the two nebulae with
  a large number of diagnostics and determine their chemical
  abundances by applying ionization correction factors
  ($\mathit{icf}$'s) to the observed ionic abundances.  For NGC~5461,
  the $\mathit{icf}$'s are based on a tailored photoionization model
  of the region (Luridiana \& Peimbert 2001), while for NGC~5471 they
  are computed from those predicted by a photoionization model of
  NGC~2363 (Luridiana, Peimbert, \& Leitherer 1999), a region which is
  similar to NGC~5471 in the ionization structure.  For both regions,
  the $\mathit{icf}$'s are compared to those computed following the
  prescriptions by Mathis \& Rosa\@ (1991).  Such comparison shows
  large discrepancies for several elements, including nitrogen, neon,
  and chlorine.  }

\begin{document}

\RescaleTitleLengths{1.07}
\maketitle
\section{INTRODUCTION}\label{sec:int}

\begin{table*}[!t]\centering 
  \setlength{\tabcolsep}{1.25\tabcolsep}
  \caption{Journal of Observations}
  \label{tab:jour_obs}
  \begin{tabular}{ccccc@{\hspace{4\tabcolsep}}cc} 
    \toprule    
    &          &        &         &                   &
    \multicolumn{2}{c}{Exposure Times (seconds)} \\
    Date       & Spectral Range & Orders &  Designation  & Slit Size        & NGC~5461           & NGC~5471    \\                                
    \midrule
    96 Jun 10& 6280--9100~\AA{}     & 25--35 & Near IR & $39.9\times 4\arcsec{}$ & $60$, $3\times1200$ & $60$, $2\times600$, $1200$ \\
    96 Jun 13& 3450--6650~\AA{}     & 34--64 & Blue          & $13.3\times 4\arcsec{}$ & $180$, $2\times900$ & $180$, $2\times600$, $900$ \\
    \bottomrule
    \\
  \end{tabular}
\end{table*} 

NGC~5461 and NGC~5471 are two giant extragalactic \ion{H}{2} regions
(GEHRs), located in the spiral galaxy M~101 (NGC~5457).  Due to their
prominence, the two GEHRs have been extensively studied by many
authors.  Low- and medium-resolution spectroscopic data in several
wavelength intervals, ranging from the IR to the UV, of one or both
regions have been published by Rayo, Peimbert, \& Torres-Peimbert\@
(1982), McCall, Rybski, \& Shields\@ (1985), Torres-Peimbert,
Peimbert, \& Fierro\@ (1989), Skillman \& Israel\@ (1988),
Casta\~neda, V\'\i lchez, \& Copetti\@ (1992), Rosa \& Benvenuti\@
(1994), and Garnett et~al.\@ (1999).  Theoretical models covering
different properties of the two regions have been described by Evans\@
(1986), Casta\~neda et~al.\@ (1992), and Luridiana \& Peimbert\@
(2001).  The chemical composition of the two regions has been
determined by Torres-Peimbert et al.\@ (1989), who aimed at
determining the chemical gradients of the parent galaxy.

In the present paper, we present high-resolution spectroscopic data of
both \ion{H}{2} regions covering a wide spectral range, from
near-ultraviolet to near-infrared.  The use of the echelle cross
disperser produces deep high-resolution spectra of a large number of
lines in the whole optical domain, overcoming the problem of
line-blending, and allowing the application of a large number of
physical-condition diagnostics.  The structure of the paper is as
follows: in \S~2 we describe the observations and the reduction
procedure; in \S~3 we present the line intensities of the two regions;
in \S~4 we derive the physical conditions in the two nebulae, and
calculate their chemical abundances by means of standard techniques;
and finally, in \S~5 we discuss our results and their implications.

\section{OBSERVATIONS AND DATA REDUCTION}\label{sec:obs}

The observations were carried out with the 2.1-m telescope of the
Observatorio Astron\'omico Nacional in San Pedro M\'artir, Baja
California, M\'exico, in June 1996.  The telescope was in its f/7.5
configuration.  High resolution CCD spectra were obtained using the
REOSC Echelle Spectrograph; see Levine \& Chakrabarty\@ (1994) for a
description of the general characteristics of this instrument.  The
echelle gives a resolution of 0.234 \AA \ pixel$^{-1}$ at H$\alpha$
using the University College of London (UCL) camera and a CCD-Tek chip
of $1024 \times 1024$ pixels with a 24 $\mu$m$^2$ pixel size. The
spectral resolution is 0.5~\AA{} FWHM and the accuracy in the
wavelength determination of emission lines is 0.1~\AA\@.

For both regions, we obtained spectra in two partially overlapped
wavelength intervals covering the range from 3450 to 9100~\AA\@.
Typically, three or four individual exposures were added to obtain the
final spectra in each interval.  Slits covering $13.3\arcsec{} \times
4\arcsec{}$ in the blue and $39.9\arcsec{} \times 4\arcsec{}$ in the
near-infrared (NIR) were used to avoid overlapping between orders in
the spatial direction. The slit orientation was east-west in all
cases. A journal of the observations is presented in
Table~\ref{tab:jour_obs}.

The atmospheric refraction depends on the observed wavelength;
therefore the region of the sky included in the slit varies with
wavelength.  This problem is called the atmospheric differential
refraction problem.  The differences in pointing caused by the
atmospheric differential refraction are not important because the
objects and the slit are relatively large. According to the slit size,
the slit orientation, the airmass ($1.09<\sec z<1.27$), and the
tables by Filippenko\@ (1982), the difference between the sampled region
by H$\beta$ and that sampled by any other line is at most 8\%; this
difference will barely affect the line intensities because the object
is considerably wider than the slit.

We used a Th-Ar lamp for wavelength calibration in all spectral ranges
and a tungsten bulb for internal flat-field images.  The absolute flux
calibration of all the spectra was achieved by taking echellograms of
the standard stars HR~4963, HR~5501 and HR~7596.  All of them are from
the list of Hamuy et al.\@ (1992), which includes bright stars with
fluxes sampled at 16~\AA{} steps. An average curve for atmospheric
extinction was used (Schuster 1982).

The spectra were reduced using the IRAF\footnote{IRAF is distributed
  by NOAO, which is operated by AURA, under cooperative agreement with
  NSF.} echelle reduction package, following the standard procedure of
bias subtraction, aperture extraction, flatfielding, wavelength
calibration and flux calibration.

\renewcommand{\multirowsetup}{\raggedleft}
\newlength{\LL} \settowidth{\LL}{3726.03}
\newlength{\LLL} \settowidth{\LLL}{26.03}
\begin{table*}\centering
  \setlength{\tabcolsep}{1.5\tabcolsep}
  \tablecols{8}
  \setlength{\tabnotewidth}{0.7\textwidth}
  \caption{Observed and Reddening-Corrected Line Ratios}
  \label{tab:line5461}
  \small
  \begin{tabular}{lr @{\hspace*{4\tabcolsep}} rrr
        @{\hspace*{4\tabcolsep}} rrr} 
    \toprule     
    & & \multicolumn{3}{c@{\hspace*{4\tabcolsep}}}{NGC~5461} & 
    \multicolumn{3}{c}{NGC~5471}\\
    Ion & 
    $\lambda_0$(\AA) & 
    $\lambda_\mathrm{obs}$(\AA) & 
    $F(\lambda)\tabnotemark{a}$ & 
    $I(\lambda)\tabnotemark{a}$ &
    $\lambda_\mathrm{obs}$(\AA) & 
    $F(\lambda)\tabnotemark{a}$ & 
    $I(\lambda)\tabnotemark{a}$ \\
    \midrule\relax
    [\ion{O}{2}]        &  3726.03 & 3728.70 & 62.00 &  85.96 
                                   & 3729.01 & 38.53 &  43.91 \\\relax

    [\ion{O}{2}]        &  3728.82 & 3731.44 & 76.16 & 105.49  
                                   & 3731.75 & 51.21 &  58.34 \\\relax

    \ion{H}{1}           &  3835.00 & 3838.22 &  5.59 & 7.48  
                                    & 3838.66 &  5.57 &   6.26 \\\relax

    [\ion{Ne}{3}]&  3868.75 & 3871.60 & 13.13 &  17.33  
                            & 3872.11 & 44.33 &  49.54 \\\relax

    \ion{He}{1} & 3888.65 & 
    \multirow{2}{\LL}{\llap{$\bigg\}$\hspace*{3\tabcolsep}}3891.71} &
    \multirow{2}{\LLL}{17.84} & \multirow{2}{\LLL}{23.48} 
    & \multirow{2}{\LL}{3892.20} &
    \multirow{2}{\LLL}{20.81} & 
    \multirow{2}{\LLL}{23.22} \\\relax

    H8 & 3889.05 & & & 
                 \\\relax


    [\ion{Ne}{3}]&  3967.47 & 3970.54 &  3.62 &   4.62 
                            & 3970.98 & 14.03 &  15.47 \\\relax

    H7                  &  3970.00 & 3973.15 & 13.62 &  17.37 
                                   & 3973.58 & 14.05 &  15.49 \\\relax

    \ion{He}{1}    &  4026.19 & 4029.50 &  1.68 &   2.12 
                              & 4029.87 &  2.04 &   2.24 \\\relax

    [\ion{S}{2}]        &  4068.60 & 4071.74 & 1.25\rlap: &  1.56\rlap: 
                                   & 4071.97 & 0.79\rlap: &  0.86\rlap: \\\relax

    [\ion{S}{2}]        &  4076.22 & 4079.30 & 0.53\rlap: &  0.66\rlap: 
                                   & 4079.72 & 0.64\rlap: &  0.70\rlap: \\\relax

    H$\delta$           &  4101.74 & 4105.03 & 23.59 &  29.09 
                                   & 4105.46 & 24.47 &  26.61 \\\relax

    H$\gamma$           &  4340.47 & 4344.11 & 47.53 &  54.67 
                                   & 4344.58 & 46.45 &  49.12 \\\relax

    [\ion{O}{3}]       &  4363.00 & 4366.76 &  1.04 &   1.19 
                                  & 4367.35 &  8.86 &   9.35 \\\relax

    \ion{He}{1}    &  4388.00 & 4391.63 & 0.33\rlap: &  0.38\rlap: 
        & \nodata & \nodata & \nodata \\

    \ion{He}{1}    &  4471.48 & 4475.26 &  4.47 &   4.98 
                              & 4475.76 &  3.96 &   4.13 \\\relax
    
    \ion{He}{2}   &  4685.68 & \nodata & \nodata & \nodata  
                             & 4690.45 &  0.38 &   0.39 \\\relax

    [\ion{Ar}{4}] &  4711.34 & \nodata & \nodata & \nodata  
                             & 4715.86 &  1.43 &   1.45 \\\relax

    [\ion{Ar}{4}] &  4740.20 & \nodata & \nodata & \nodata  
                             & 4745.10 &  1.17 &   1.18 \\\relax

    {H$\beta$}          &  4861.33 & 4865.54 &100.00 & 100.00 
                                   & 4866.05 &100.00 & 100.00 \\\relax

    \ion{He}{1}    &  4921.93 & 4925.97 &  1.50 &   1.48 
                              & 4926.85 &  0.95 &   0.95 \\\relax

    [\ion{O}{3}]       &  4958.92 & 4963.28 &114.52 & 112.17 
                                  & 4963.80 &232.86 & 230.94 \\\relax

    [\ion{O}{3}]       &  5006.85 & 5011.15 &333.51 & 323.98 
                                  & 5011.70 &684.10 & 676.22 \\\relax

    [\ion{Cl}{3}]&  5517.45 & 5522.52 & 0.54\rlap: &  0.47\rlap: 
                            & 5522.92 & 0.33\rlap: &  0.31\rlap: \\\relax

    [\ion{N}{2}]        &  5754.57 & 5759.55 &  0.48 &   0.40 
                                   & 5759.45 & 0.35\rlap: &  0.32\rlap: \\\relax

    \ion{He}{1}    &  5875.67 & 5880.65 & 16.62 &  13.37 
                              & 5881.28 & 12.35 &  11.32 \\\relax    

    [\ion{O}{1}]         &  6300.32 & 6306.11 &  5.57 &  4.13 
                                    & 6306.34 &  4.41 &  3.92 \\\relax 

    [\ion{S}{3}]       &  6312.06 & 6317.36 &  1.76 &  1.29 
                                  & 6317.90 &  1.82 &  1.62 \\\relax

    [\ion{O}{1}]        &  6363.81 & 6369.17 &  0.97 &  0.71 
                                   & 6369.69 &  0.90 &  0.80 \\\relax

    [\ion{N}{2}]        &  6548.03 & 6553.59 & 21.02 & 14.90 
        & \nodata & \nodata & \nodata \\

    H$\alpha$           &  6562.82 & 6567.83 &408.31 &288.57 
                                   & 6568.54 &321.83 &280.11 \\\relax

    [\ion{N}{2}]        &  6583.41 & 6588.93 & 59.33 & 41.71 
                                   & 6589.45 &  8.62 &  7.48 \\\relax

    \ion{He}{1}    &  6678.15 & 6684.01 &  4.80 &  3.31 
                              & 6684.60 &  3.73 &  3.21 \\\relax

    [\ion{S}{2}]        &  6716.47 & 6722.27 & 29.42 & 20.07 
                                   & 6722.74 & 12.70 & 10.90 \\\relax

    [\ion{S}{2}]        &  6732.07 & 6736.64 & 22.90 & 15.59 
                                   & 6737.12 &  9.44 &  8.09 \\\relax

    \ion{He}{1}    &  7065.28 & 7070.84 &  3.22 &  2.13 
                              & 7071.71 &  2.85 &  2.41 \\\relax

    [\ion{Ar}{3}]&  7135.93 & 7141.91 & 15.52 & 10.12 
                            & 7142.49 &  7.91 &  6.68 \\\relax

    [\ion{O}{2}]        &  7319.65 & 7326.60 &  3.49 &  2.22 
                                   & 7327.25 &  1.99 &  1.66 \\\relax

    [\ion{O}{2}]        &  7330.16 & 7337.01 &  3.03 &  1.93  
                                   & 7337.26 &  1.96 &  1.63 \\\relax

    [\ion{S}{3}]       &  9068.90 & 9076.42 & 53.34 & 28.47 
                                  & 9077.07 & 24.23 & 18.84 \\
    \midrule

    \multicolumn{2}{l}{$\log F(\mathrm{H}\beta)$\tabnotemark{b}} 
    &\multicolumn{3}{c@{\hspace*{4\tabcolsep}}}{$-12.41$}     
    &\multicolumn{3}{c}{$-12.53$}     \\

    \multicolumn{2}{l}{$\log \mathrm{EW(H\beta)}$\tabnotemark{c}}
    &\multicolumn{3}{c@{\hspace*{4\tabcolsep}}}{$2.18$}       
    &\multicolumn{3}{c}{$2.13$}       \\

    \multicolumn{2}{l}{$C(\mathrm{H}\beta$)}         &\multicolumn{3}{c@{\hspace*{4\tabcolsep}}}{$0.45\pm0.06$}
                                              &\multicolumn{3}{c}{$0.18\pm0.06$}\\

    \bottomrule
    \tabnotetext{a}{Normalized to  H$\beta=100.00$.\quad 
      \textsuperscript{b}Units of erg sec$^{-1}$ cm$^{-2}$.\quad
      \textsuperscript{c}Units of \AA\@.}
  \end{tabular}
\end{table*}

\section{THE EMISSION-LINE SPECTRA}\label{sec:dat}

The line fluxes and the dereddened intensities for NGC~5461 and
NGC~5471, normalized to $I(\mathrm{H}\beta)=100$, are listed in
Table~\ref{tab:line5461}.  The line fluxes were measured with the
\texttt{splot} task of IRAF\@.  The central wavelengths were measured
with \texttt{splot} and referred to the heliocentric reference frame.
The observational uncertainties associated with the line flux
intensities are estimated to be 0.02~dex for those lines with
$F(\lambda)/F(\mathrm{H}\beta) \geq 0.10$, 0.04~dex for those lines with
$0.02 \leq F(\lambda)/F(H\beta) < 0.10$, and 0.08~dex for those lines
with $F(\lambda)/F(\mathrm{H}\beta) < 0.02$.  Those lines marked with a
colon are affected by an uncertainty of about 0.15~dex.

Since the NIR and the blue spectra have been taken with different
apertures, it is advisable to correct for this effect (as well as for
possible shifts in the telescope position from one night to another)
to compute line ratios in a homogeneous manner.  For this reason, we
applied a grey shift to the line intensities of the NIR spectra lines,
by matching the H$\alpha$ intensities.  The intensity of the blue
spectrum has been considered for those lines falling in both
wavelength ranges ([\ion{O}{1}]~$\lambda\lambda$~6300, 6363,
[\ion{N}{2}]~$\lambda\lambda$~6548, 6584, [\ion{S}{3}]~$\lambda$~6312,
and H$\alpha$).  The shift applied was $+0.032$~dex for NGC~5461 and
$-0.109$~dex for NGC~5471.  Note that for NGC~5461
$I(\mathrm{H}\alpha)_\mathrm{blue}/I(\mathrm{H}\alpha)_\mathrm{NIR} >
1$, even though the slit used in the blue range was smaller than the
one used in the NIR: this fact probably implies that the telescope was
not pointing exactly to the same position during the two nights, an
effect amplified by the extremely-peaked brightness distribution of
the region. Indeed, Figure 13 in Casta\~neda et al.\@ (1992) and
Figure 3 in Luridiana \& Peimbert (2001) show that the H$\alpha$
intensity falls to less than 10\% of the peak value within 3$''$ from
the center.

The NIR lines were corrected for telluric absorption; this correction
was smaller than 3\% for all the lines with the exception of
$\lambda\lambda 7320,\, 9069$ in NGC 5461 where the correction
amounted to 6\% and 7\%, and of $\lambda 9069$ in NGC 5471 where the
correction amounted to 9\%.  The reddening coefficient, $C$(H$\beta$),
was determined comparing the
$\mathrm{\mathnormal{I}(H\beta)/\mathnormal{I}(H\alpha)_{blue}}$
observed ratio to the case B one computed by Storey \& Hummer\@ (1995)
adopting $T_\mathrm{e}=8700$~K and $N_\mathrm{e}=150$~cm$^{-3}$ for
NGC~5461, and $T_\mathrm{e}=12,900$~K, $N_\mathrm{e}=90$~cm$^{-3}$ for
NGC~5471.  The uncertainty in $C(\mathrm{H}\beta)$ has been estimated
assuming a $5\%$ uncertainty in $I(\mathrm{H}\beta)$ and
$I(\mathrm{H\alpha)_{blue}}$.  The results are shown in
Table~\ref{tab:line5461}, together with the total uncorrected H$\beta$
fluxes, measured in erg s$^{-1}$ cm$^{-2}$, the equivalent widths of
H$\beta$, and the reddening coefficients.

\section{PHYSICAL CONDITIONS AND CHEMICAL ABUNDANCES}\label{sec:ana}

The physical conditions in the two nebulae have been calculated with
the \texttt{ntplot} and the \texttt{temden} IRAF (version 2.11.3)
tasks.  These tasks are based on 5-level model atoms for [\ion{O}{2}],
[\ion{S}{3}], [\ion{Cl}{3}], and [\ion{Ar}{4}], 6-level model atoms
for [\ion{O}{3}] and [\ion{N}{2}], and an 8-level model atom for
[\ion{S}{2}].  The \texttt{ntplot} task allows a self-consistent
computation of $N_\mathrm{e}$ and $T_\mathrm{e}$ for a given ion, when
the relevant line ratios are known; it was used to determine
$N_\mathrm{e}(\ion{O}{2})$, $T_\mathrm{e}(\ion{O}{2})$, and
$N_\mathrm{e}(\ion{S}{2})$, $T_\mathrm{e}(\ion{S}{2})$ for both
\ion{H}{2} regions.  When only one line ratio is known for a given
ion, the task \texttt{temden} is used, which computes $T_\mathrm{e}$
for an assumed $N_\mathrm{e}$ value or vice versa.
$T_\mathrm{e}(\ion{N}{2})$, $T_\mathrm{e}(\ion{O}{3})$, and
$T_\mathrm{e}(\ion{S}{3})$ were calculated with \texttt{temden}
assuming $N_\mathrm{e} = N_\mathrm{e}(\ion{O}{2})$, and
$N_\mathrm{e}(\ion{Ar}{4})$ was calculated assuming
$T_\mathrm{e}=T_\mathrm{e}(\ion{O}{3})$.  It is worthwhile to remark
that the high resolution and wide wavelength range of these
observations considerably extends the number of available
physical-condition diagnostics of the two \ion{H}{2} regions, adding
$N_\mathrm{e}(\ion{O}{2})$, $N_\mathrm{e}(\ion{Ar}{4})$ and
$T_\mathrm{e}(\ion{S}{3})$ to the diagnostics studied in previous
works.

\newcommand{\DS}{\hspace{6\tabcolsep}} 
\newcommand{\HS}{\hspace{3\tabcolsep}} 
\newcommand{\LS}{\hspace{2pt}} 
\newcommand{\updown}[2]{%
  \rlap{\raisebox{0.8ex}{\scriptsize #1}}%
  \raisebox{-0.5ex}{{\scriptsize #2}}}
\begin{table*}\centering
  \setlength{\tabcolsep}{1.6\tabcolsep}
  \setlength{\tabnotewidth}{0.9\textwidth}
  \setlength{\defaultaddspace}{0.5ex}
  \tablecols{13}
  \caption{Electron Temperatures and Densities of NGC~5461
    and NGC~5471} \label{tab:phyc}
  \begin{tabular}{
      l 
      @{\DS} r@{\LS}c@{\LS}l r@{\LS}c@{\LS}l 
      @{\DS} r@{\LS}c@{\LS}l r@{\LS}c@{\LS}l
      @{\HS}
      }
    \toprule

    Quantity\tabnotemark{a} & \multicolumn{6}{c@{\DS}}{NGC~5461} &
    \multicolumn{6}{c@{\HS}}{NGC~5471} \\ 
    \midrule

    $T_\mathrm{e}$(\ion{O}{2})  & 10,400&$\pm$& 500 & 9500 &$\pm$&  900
    & 14,200 &$\pm$&  900               & 13,100 &$\pm$& 2000    \\ 

    $T_\mathrm{e}$(\ion{S}{2})  & 8000 &$\pm$&1600 & 9500 &$\pm$& 1100
    & 10,000 &$\pm$& 3200               & 12,700 &$\pm$& 2400    \\ 

    $T_\mathrm{e}$(\ion{N}{2})  & 8700 &$\pm$& 700 & 8500 &$\pm$&  550
    & 10,800 &$\pm$& 2300               & 10,800 &$\pm$& 1700    \\ 

    $T_\mathrm{e}$(\ion{O}{3}) & 8500 &$\pm$& 500 & 9300 &$\pm$&  250
    & 13,000 &$\pm$&  500               & 13,400 &$\pm$&  250    \\

    $T_\mathrm{e}$(\ion{S}{3}) & 9000 &$\pm$& 800 &
    \multicolumn{3}{c@{\DS}}{\nodata}      & 12,100 &$\pm$&  800
    &  \multicolumn{3}{c@{\HS}}{\nodata}         \\ 

    \addlinespace

    $N_\mathrm{e}$(\ion{O}{2})  & 150  &$\pm$&  60 &
    \multicolumn{3}{c@{\DS}}{\nodata}      &    90 &$\pm$& 70
    &  \multicolumn{3}{c@{\HS}}{\nodata}         \\  

    $N_\mathrm{e}$(\ion{S}{2})  & 130  &$\pm$&  90 &
    \multicolumn{3}{c@{\DS}}{234}   &    70 &$\pm$&$\updown{100}{70}$  &
    \multicolumn{3}{c@{\HS}}{186}       \\ 

    $N_\mathrm{e}$(\ion{Ar}{4}) & \multicolumn{3}{c}{\nodata}     &
    \multicolumn{3}{c@{\DS}}{\nodata}      &  1350
    &$\pm$&$\updown{1150}{920}$&  \multicolumn{3}{c@{\HS}}{\nodata}
    \\  

    $N_\mathrm{e}$(rms)         & \multicolumn{3}{c}{14.2}           &
    \multicolumn{3}{c@{\DS}}{14.8} & \multicolumn{3}{c}{10.4}  &
    \multicolumn{3}{c@{\HS}}{10.6}        \\ 

    \addlinespace

    References         &   \multicolumn{3}{c}{1} &
    \multicolumn{3}{c@{\DS}}{2}  &   \multicolumn{3}{c}{1}        &
    \multicolumn{3}{c@{\HS}}{2}     \\

    \bottomrule

    \tabnotetext{a}{Given in K and cm$^{-3}$, respectively.\qquad 
      \textsc{References}:--- (1) This work; (2) Torres-Peimbert et al.\@ 1989.}
  \end{tabular}
\end{table*}

Table~\ref{tab:phyc} lists the physical conditions of the two nebulae.
The results obtained are in good agreement with those by
Torres-Peimbert et al.\@ (1989) within the estimated errors.  The only
notable exception is $T_\mathrm{e}(\ion{S}{2})$, for which
Torres-Peimbert et~al.\@ (1989) found much higher values in both
objects.  However, such differences depend on the atomic parameters
used, rather than on differences in the measured line ratios: indeed,
using the intensity values of Torres-Peimbert et al.\@ (1989) and the
same atomic parameters as us, one would obtain slightly lower
temperatures than ours (the atomic parameters used in this paper are
those from Verner, Verner, \& Ferland 1996, Keenan et al. 1993, and
Ramsbottom, Bell, \& Stafford 1996).

The root mean square density, $N_\mathrm{e}(\mathrm{rms})$, to a very
good approximation is given by the expression: \beq
N_\mathrm{e}(\mathrm{rms}) = \left\{\frac{ I(\mathrm{H}\beta) \> [1 +
    N(\mathrm{He^+})/N(\mathrm{H^+})] \> 4\pi d^2}{V \>
    \alpha(\mathrm{H}\beta, T_\mathrm{e}) \>
    h\nu_{\scriptscriptstyle\mathrm{H}\beta}}\right\}^{1/2}, \eeq
where $V$ is the observed volume, $\alpha(\mathrm{H\beta},
T_\mathrm{e})$ is the H$\beta$ effective recombination coefficient,
and $d$ is the distance to the region.  We assume $V = 4\pi r^3/3$,
where $r$ is the radius of a circle equivalent in area to our slit, at
a distance of 7.4 Mpc (Sandage \& Tammann 1976, with the suggested
correction by de Vaucouleurs 1978).  The results, substantially
equivalent to those of Torres-Peimbert et al.\@ (1989), are also
listed in Table~\ref{tab:phyc}.  {From} these values, combined with
$N_\mathrm{e}$(\ion{O}{2}) or $N_\mathrm{e}$(\ion{S}{2}), we can
calculate the filling factor, which is $0.009\lesssim \epsilon
\lesssim 0.011$ for NGC~5461 and $0.013\lesssim \epsilon \lesssim
0.022$ for NGC~5471.  These values can be compared to the filling
factor of the composite model of NGC~5461 by Luridiana \& Peimbert
(2001), which averages 0.003.  The lower filling factor derived by
Luridiana \& Peimbert (2001) was based on the data by Torres-Peimbert
et al.\@ (1989), which obtained a higher $N_\mathrm{e}(\ion{S}{2})$
than that derived here.

Small filling factors of $\sim0.01$ are characteristic of GEHRs
(Shields 1986). Similar values of about 0.02 have been found in the
ionized gas of emission-line dwarf galaxies (Martin 1997).  In
contrast, Copetti et~al.\@ (2000) find a mean filling factor of 0.1
for a sample of 15 Galactic ``normal'' \ion{H}{2} regions. Although
the estimations of $\epsilon$ in ionized nebulae are indirect and very
rough, the difference between ``normal'' and giant \ion{H}{2}
regions---if real---would imply a higher clumpiness in the giant ones.
This is not unexpected considering the typical highly disturbed
morphology and complex kinematics of GEHRs, that may be produced by
the dynamical effects of stellar winds and supernovae in GEHRs.

\subsection{Ionic Abundances}

\subsubsection{Heavy Elements}\label{sec:heavies}

The ionic abundances of heavy elements in the two regions have been
calculated with the IRAF task \texttt{abund}, which is based on a
three-zone model of the nebula, with the three zones corresponding to
the low-, medium-, and high-ionization regions, each characterized by
representative $N_\mathrm{e}$ and $T_\mathrm{e}$ values.

Concerning the temperature, each of the three low-ionization
diagnostics available---\tnii, \tsii, and \toii---is affected by
uncertainties: (a)~Nitrogen lines are weak in these objects,
especially in NGC~5471.  (b)~\toii{} depends non-negligibly on the
assumed reddening, and on the assumed density; furthermore,
\oii~\lloii{} are weak and affected by telluric absorption lines.
Finally, (c)~\sii~\llsii{} are also weak, and the estimates of the
\sii{} atomic parameters are constantly changing; additionally,
\tsii{} has a dependence on the assumed \eden{} similar to that of
\toii.  To minimize these uncertainties, we decided to adopt the
average temperature of the three diagnostics, i.e., $\etem = 9000$~K
for NGC~5461 and $\etem = 11,700$~K for NGC~5471.  Finally, in both
the medium- and high-ionization zones we adopted \toiii{} as a
representative temperature in both regions, since \tsiii{} is more
uncertain due to the observational errors in the sulfur lines and the
frequent revisions of the \siii{} atomic parameters.

Concerning the electronic density, we adopted in the three zones of
both \ion{H}{2} regions $N_\mathrm{e} = N_\mathrm{e}(\ion{O}{2})$.  We
preferred $N_\mathrm{e}(\ion{O}{2})$ over $N_\mathrm{e}(\ion{S}{2})$,
due to the following facts: (a)~the \oii{} lines are more intense than
the \sii{} lines, (b)~the \oii{} electronic density diagnostic is
slightly more sensitive than the \sii{} one in the low-density regime,
and (c)~the \sii{} atomic parameters are still highly uncertain.

With these assumptions, we have obtained the abundances of N$^+$,
O$^0$, O$^+$, O$^{++}$, Ne$^{++}$, S$^+$, S$^{++}$, Cl$^{++}$,
Ar$^{++}$, and Ar$^{3+}$ (Table~\ref{tab:ion_ab}).

\begin{table}\centering
  \setlength{\tabnotewidth}{0.6\columnwidth}
  \tablecols{3}
  \setlength{\tabcolsep}{3\tabcolsep}
  \caption{Ionic abundances\tabnotemark{a} in 
    NGC~5461 and NGC~5471 } \label{tab:ion_ab}
  \begin{tabular}{lrr}
    \toprule
    Ion & \multicolumn{1}{c}{NGC~5461} & \multicolumn{1}{c}{NGC~5471} \\
    \midrule
    O$^0$    & $7.08\pm0.20$ & $6.63\pm0.20$\\
    O$^+$    & $8.08\pm0.14$ & $7.32\pm0.14$\\
    O$^{++}$ & $8.32\pm0.07$ & $8.02\pm0.07$\\
    N$^+$    & $7.04\pm0.12$ & $6.01\pm0.13$\\
    Ne$^{++}$ & $7.59\pm0.11$ & $7.32\pm0.10$\\
    S$^+$    & $6.02\pm0.19$ & $5.47\pm0.20$\\
    S$^{++}$ & $7.00\pm0.10$ & $6.45\pm0.10$\\
    Cl$^{++}$ & $4.93\pm0.16$ & $4.20\pm0.16$\\
    Ar$^{++}$ & $6.15\pm0.12$ & $5.55\pm0.14$\\
    Ar$^{3+}$ &\multicolumn{1}{c}{\nodata} & $5.07\pm0.10$\\
    \bottomrule
    \tabnotetext{a}{Given as $12 + \log ( \mathrm{X}^{+i}/\mathrm{H}^+ )$.}
  \end{tabular}
\end{table}

\subsubsection{Helium}

There are 8 lines of the \ion{He}{1} recombination spectrum observed
in NGC~5461 and 7 lines in NGC~5471.  To calculate the He$^+$
abundance, it is necessary to correct the line intensities for the two
well-known effects of self-absorption and collisional enhancement.
The correction for self-absorption may be rather uncertain for
$\lambda~3889$ and $\lambda~7065$; furthermore the observational
errors in the line intensities of $\lambda\lambda$~4026, 4388, and
4922 are relatively large.  Consequently we decided to determine the
He$^+$ abundances based on the $\lambda\lambda$~4471, 5876 and 6678
lines.

The He$^+$ abundance can be computed for each \ion{He}{1} line by
means of the relation: \beq y^+ \equiv \mathrm{\frac{He^+}{H^+}} =
\frac{\lambda}{4861} \>
\mathrm{\frac{\alpha_{eff}(H\beta)}{\alpha_{eff}(\lambda)}} \>
\frac{I(\lambda)^\mathrm{R}}{I(\mathrm{H}\beta)}, \eeq where
$\alpha_\mathrm{eff}(\lambda)$ is the effective recombination
coefficient of the considered transition, and $I(\lambda)^\mathrm{R}$
is the observed intensity of the considered line, corrected for
collisional enhancement and optical depth effects.

The collisional contribution to the helium lines was estimated from
Kingdon \& Ferland (1995) and Banjamin, Skillman, \& Smits (1999)
assuming $\etem{} = 8700$~K and $\eden{} = 150$~cm$^{-3}$ for
NGC~5461, and $\etem{} = 12,900$~K and $\eden{} = 90$~cm$^{-3}$ for
NGC~5471.  We decided to adopt for the temperature a weighted average
of $T(\ion{O}{3})$ as representative of the O$^{++}$ region and an
average of $T(\ion{O}{2})$, $T(\ion{S}{2})$ and $T(\ion{N}{2})$ as
representative of the O$^+$ region; since the emissivity is dominated
by the O$^{++}$ region the adopted temperatures are closer to the
$T(\ion{O}{3})$ values.  The self absorption effects in the triplet
lines were estimated from the computations by Robbins (1968), adopting
$\tau(3889) = 1.0$ and 0.2 for NGC~5461 and NGC~5471, respectively.
The $\tau(3889)$ values were estimated from the $\lambda\lambda$~3889
and 7065 line intensities and from the photoionization model of
NGC~5461 by Luridiana \& Peimbert (2001) computed with {\sc Cloudy}
(Ferland 1996).  The $\alpha_\mathrm{eff}(\lambda)$ coefficients at
$\etem{} = 8700$~K (NGC~5461) and $\etem{} = 12,900$~K (NGC~5471) have
been obtained by power-law interpolation in the appropriate
temperature interval between the data provided by Smits (1996) for
$N_\mathrm{e}=100$~cm$^{-3}$.  Given the low densities of the two
\ion{H}{2} regions, the correction for collisional enhancement is
generally small; similarly, the correction for optical depth effects
is very small because $\tau(3889)$ is relatively small.  Finally, the
effective recombination coefficient for H$\beta$ is based on the data
by Brocklehurst (1971) for $N_\mathrm{e}=100$~cm$^{-3}$, assuming a
power-law interpolation for $\etem{} = 8700$~K (NGC~5461) and $\etem{}
= 12,900$~K (NGC~5471).

Table~\ref{tab:helium} presents the resulting $y^+$ values, together with the 
mean value. The mean value was obtained by weighting the different lines as
the square root of their intensities.

\begin{table}\centering
  \setlength{\tabcolsep}{2.6\tabcolsep}
  \setlength{\defaultaddspace}{0.3ex}
  \caption{Helium abundance values}\label{tab:helium}
  \begin{tabular}{lc@{\qquad}c}
    \toprule
    & \multicolumn{2}{c}{$y^+$} \\ \cmidrule(lr){2-3}
    Line &  NGC~5461 & NGC~5471\\
    \midrule
    4471 & 0.0983  & 0.0846  \\
    5876 & 0.0945  & 0.0867  \\
    6678 & 0.0836  & 0.0879  \\
    \addlinespace
    Mean value & 0.0937 & 0.0864 \\
    \bottomrule
  \end{tabular}
\end{table}

There are three minor effects that were not considered in the helium
abundance determination: (a)~the correction due to the temperature
structure, (b)~the correction due to the collisional excitation of the
Balmer lines and (c)~the correction for the underlying absorption of
the helium lines. To estimate the first two effects observations of
higher accuracy are required (e.g., Peimbert, Peimbert, \& Ruiz 2000;
Peimbert, Peimbert, \& Luridiana 2002; Luridiana, Peimbert, \&
Peimbert 2002), the third effect is considered by Esteban et~al.\@
(2002) for these objects.

In NGC~5471, \ion{He}{2}~$\lambda~4686$ is also observed, implying the
presence of He$^{++}$. The abundance of He$^{++}$ relative to H$^{+}$
can be computed by means of the equation: \beq
\mathrm{\frac{He^{++}}{H^+}} = \frac{4686}{4861} \>
\mathrm{\frac{\alpha_{eff}(H\beta)}{\alpha_{eff}(4686)}} \>
\frac{I(4686)}{I(\mathrm{H}\beta)}, \eeq where the
$\alpha_\mathrm{eff}(\lambda)$ coefficients at $\etem{} = 12900$~K
have been obtained by a power-law interpolation in temperature from
the data provided by Smits\@ (1996) for $N_\mathrm{e}=100$~cm$^{-3}$.
The result of this calculation is $\mathrm{He^{++}/H^+} = 3.9 \times
10^{-4}$, a completely negligible quantity from the point of view of
the total abundance.  It is interesting to note that in our spectra
the nebular \ion{He}{2}~$\lambda~4686$ line is superposed on a much
broader stellar line, about 3.5 times more intense than the nebular
line; the sum of the two components is roughly a factor of 1.7 higher
than the $I(\lambda 4686)$ reported by Torres-Peimbert et al.\@
(1989), while the nebular line alone is about one third of it.  These
differences are probably a combination of aperture effects, and the
much lower resolution used by Torres-Peimbert et~al.\@ (1989).

The broad component of $\lambda~4686$ is due to WR stars; in addition
we also observe stellar features of \ion{N}{5}~4602, 4620,
\ion{C}{4}~4658, and \ion{C}{4}~5808, whereas \ion{N}{3}~4640 and
\ion{C}{3}~4650 are not detected (see Table~\ref{tab:stellar_lines}).
All the detected lines carry uncertainties of the order of 20\%.  The
\ion{N}{5}~4602, 4620 feature implies the presence of WN stars, while
the two \ion{C}{4} features implies the presence of WC or WO stars;
the low 4658/5808 intensity reatio, compared to the observed ratios
compiled by Schaerer \& Vacca\@ (1998), might suggest that the WO
stars dominate and the WC stars are not numerous or even absent.
However, it is not possible to interpret quantitatively the observed
WR features in terms of a distribution among the various WR spectral
types, due to the uncertainties both in our observations and in the
calibration of WR spectra.  Nevertheless, we can give a rough estimate
of the \emph{total} number of WR stars, using average values for the
``WR bump'' luminosities for WN and WC stars (Smith 1991; Schaerer \&
Vacca 1998): We obtain in this way 14 ``equivalent'' WC4 stars and 26
``equivalent'' WN7 stars; the WC4 stars account for all the 5808~\AA{}
bump flux, and almost one half of the 
$\ion{N}{iii/v} + \ion{C}{iii/iv} + \ion{He}{ii}$
blend around 4650~\AA{}, and the WN7 stars for the rest of the flux in
the 
$\ion{N}{iii/v} + \ion{C}{iii/iv} + \ion{He}{ii}$
blend.  On the other hand, the observed H$\beta$ flux corresponds to a
number of ionizing photons $Q(\mathrm{H}^0)\sim 5 \times
10^{51}$~s$^{-1}$ (assuming 0.8 for the covering factor), or 450
equivalent O7V stars (Vacca 1994) ; for a Salpeter's IMF with
$M_\mathrm{up}= 120~M_\odot$, this corresponds to a total number of about
530 O~stars in the region (Schaerer \& Vacca 1998).  From these
estimates we derive a WR/O ratio of $40/530 \sim 0.075$.

\begin{table}[!t]\centering
  \tablecols{5}
  \setlength{\tabnotewidth}{\columnwidth}
  \setlength\tabcolsep{1.2\tabcolsep}
  \caption{Stellar lines of NGC 5471 } \label{tab:stellar_lines}
  \begin{tabular}{llccc}
    \toprule
    \multicolumn{1}{c}{$\lambda$ (\AA)} & \multicolumn{1}{c}{Ion} &
    $F(\lambda)$\tabnotemark{a} & $I(\lambda)$\tabnotemark{a} &
    $L(\lambda)$\tabnotemark{b} \\
    \midrule
    $4604+4620$ & \ion{N}{v}   & 5.6 & 9.0 & 5.9 \\
    4658        & \ion{C}{iv}  & 4.9 & 7.8 & 5.1 \\
    4686        & \ion{He}{ii} & 3.9 & 6.3 & 4.1 \\
    $5801+5812$ & \ion{C}{iv}  & 4.5 & 6.5 & 4.3 \\
    \bottomrule
    \tabnotetext{a}{Units: $10^{-15}$ erg cm$^{-2}$ s$^{-1}$.\quad
      \textsuperscript{b}Units: $10^{+37}$ erg s$^{-1}$.}
  \end{tabular}
\end{table}

\subsection{Chemical Abundances}

The chemical abundances for both \ion{H}{2} regions were calculated
under the assumption of no temperature fluctuations, and without
considering the fraction of heavy elements trapped in dust.  The
resulting values are listed in Table~\ref{tab:chem_ab}, which also
contains the gaseous abundances derived by Torres-Peimbert et al.\@
(1989) under the assumption that $t^2=0$, and the Orion and solar
chemical abundances compiled from the literature.  For both objects
the $Y$ values derived by us are higher than those derived by
Torres-Peimbert et~al.\@ (1989): we prefer our values due to the
higher resolution of our spectra.  For NGC~5461 the N/O and Ne/O
values derived by us are higher than those of Torres-Peimbert et al.\@
(1989): the difference is mostly due to differences in the
$\mathit{icf}$'s used by the two groups.  For NGC~5471 the N/O and
Ne/O values derived by us are similar to those by Torres-Peimbert
et~al.\@ (1989), but in this case the $\mathit{icf}$'s used were
almost the same.  The gaseous values for the heavy species in Orion
are taken from Esteban et~al.\@ (1998), with the exception of argon
that was taken from Peimbert (1993); the solar values are taken from
Holweger (2001) (N and Ne), Grevesse \& Sauval (1998) (S, Cl, and Ar),
and from Christensen-Dalsgaard\@ (1998) (He), while the oxygen
abundance is an average of the values cited in Allende-Prieto,
Lambert, \& Asplund (2001) and Holweger (2001).

\newcommand{\SSS}{\hspace{\tabcolsep}} 
\begin{table*}[!t]
  \setlength{\tabnotewidth}{\textwidth}
  \setlength\tabcolsep{1.2\tabcolsep}
  \tablecols{13}
  \setlength{\defaultaddspace}{0.3ex}
  \caption{Gaseous chemical abundances in NGC~5461, NGC~5471, Orion, and the Sun}
  \label{tab:chem_ab}
  \centering
  \small
  \newlength{\RefWidth} \settowidth{\RefWidth}{\textsc{References}:---
    }
  \begin{tabular}{
      l
      r@{\LS$\pm$\LS}l
      @{\SSS}
      r@{\LS$\pm$\LS}l
      r@{\LS$\pm$\LS}l
      @{\SSS}
      r@{\LS$\pm$\LS}l
      r@{\LS$\pm$\LS}l
      r@{\LS$\pm$\LS}l}
    \toprule
    &\multicolumn{4}{c}{NGC~5461}  &\multicolumn{4}{c}{NGC~5471} &\multicolumn{2}{c}{Orion}       &\multicolumn{2}{c}{Sun}\\
    Element         &\multicolumn{4}{c}{($t^2=0$)}&\multicolumn{4}{c}{($t^2=0$)}&\multicolumn{2}{c}{($t^2=0.024$)}&\multicolumn{2}{c}{}\\
    \midrule
    $Y$            & $0.285$&$0.011$& $0.264$&$0.007$           & $0.257$&$0.010$& $0.241$&$0.007$& $0.276$&$0.007$& $0.271$&$0.010$\\
    $12 + \log \mathrm{O/H}$    &  $8.52$&$0.10$ &  $8.39$&$0.08$            &  $8.10$&$0.08$ &  $8.05$&$0.05$ &  $8.64$&$0.06$ &  $8.71$&$0.05$ \\
    $\log \mathrm{N/O}$         & $-0.64$&$0.12$ & $-1.13$&$0.06$            & $-1.23$&$0.10$ & $-1.33$&$0.06$ & $-0.85$&$0.10$ & $-0.78$&$0.12$ \\
    $\log \mathrm{Ne/O}$        & $-0.44$&$0.12$ & $-0.68$&$0.06$            & $-0.73$&$0.10$ & $-0.63$&$0.06$ & $-0.74$&$0.12$ & $-0.71$&$0.09$ \\
    $\log \mathrm{S/O}$         & $-1.46$&$0.11$ & $-1.69$&$0.08$            & $-1.57$&$0.10$ & $-1.67$&$0.08$ & $-1.46$&$0.12$ & $-1.38$&$0.12$ \\
    $\log \mathrm{Cl/O}$        & $-3.54$&$0.20$ &\multicolumn{2}{c}{\nodata}& $-3.67$&$0.20$ & $-3.30$&$0.16$ & $-3.21$&$0.03$ & $-3.21$&$0.30$ \\
    $\log \mathrm{Ar/O}$        & $-2.35$&$0.12$ & $-2.21$&$0.08$            & $-2.38$&$0.15$ & $-2.25$&$0.08$ & $-2.15$&$0.21$ & $-2.31$&$0.08$ \\
    \addlinespace
    References      & \multicolumn{2}{c@{\SSS}}{1} &
    \multicolumn{2}{c}{2} & \multicolumn{2}{c@{\SSS}}{1}
    & \multicolumn{2}{c}{2} & \multicolumn{2}{c}{3, 4}
    & \multicolumn{2}{c}{5, 6, 7, 8 } \\
    \bottomrule
    \multicolumn{13}{@{}p{\textwidth}@{}}{\hangindent\RefWidth
      \textsc{References}:--- (1)~This work; (2)~Torres-Peimbert 
      et~al.\@ 1989; (3)~Esteban et~al. 1998; (4)~Peimbert 1993; 
      (5)~Allende-Prieto et~al.\@ 2001; (6)~Holweger 2001;
      (7)~Christensen-Dalsgaard 1998; (8)~Grevesse \& Sauval 1998.}
  \end{tabular}
\end{table*}

\subsubsection{NGC~5461}

The computation of the chemical abundances from the observed ionic
abundances requires a knowledge of the ionization structure of the
model.  In the case of NGC~5461, we can rely on the ionization
structure predicted by the photoionization model of Luridiana \&
Peimbert (2001 and private communication).  Radial and volumetric
integrations of the ionic fractions are presented in
Table~\ref{tab:ioniz_av}.  The radial average for a given ion
$\mathrm{X}^i$ is defined as 
\begin{equation}
  \left(\frac{{\mathrm{X}}^i}{\mathrm{X}}\right)_{\!\!R}
  \equiv 
  \int_0^{R_\mathrm{S}} \kern-1.6ex N({\mathrm{X}}^i) \> N_\mathrm{e} \> dR
  \,\Bigg/\!
  \int_0^{R_\mathrm{S}} \kern-1.6ex N({\mathrm{X}}) \>  N_\mathrm{e} \> dR, 
\end{equation}
while the volume average is: 
\begin{equation}
  \left(\frac{{\mathrm{X}}^i}{\mathrm{X}}\right)_{\!\!V}
  \equiv 
  \int_V  N({\mathrm{X}}^i) \> N_\mathrm{e} \> dV
  \,\Bigg/\!
  \int_V N({\mathrm{X}}) \> N_\mathrm{e} \> dV.
\end{equation}

{From} these average ionic ratios, the ionization correction factors can be
easily computed, e.g., for the radial $\mathit{icf}$'s:
\begin{equation}
  \label{eq:1}
  \mathit{icf}({\mathrm{X}}^i)^{\mathrm{LP01}, R} \equiv 
  \frac{({\mathrm{X}}/{\mathrm{X}}^{i})}{({\mathrm{H}}/{\mathrm{H}^+})} = 
  \frac{({\mathrm{H}^+}/{\mathrm{H}})_R}{({\mathrm{X}}^i/{\mathrm{X}})_R},
\end{equation}
and analogously for the volume-weighted ionization correction factors.

\begin{table*}\centering
  \setlength{\tabnotewidth}{0.40\textwidth}
  \setlength{\tabcolsep}{1.33\tabcolsep}
  \tablecols{10}
  \caption{Ionization fractions from the photoionization model
    of NGC~5461\tabnotemark{a}} \label{tab:ioniz_av} 
  \begin{tabular}{l @{\DS} cccc l cccc}
    \toprule
    & \multicolumn{9}{c}{Ionization Stage}\\
    \cmidrule{3-9}
    & \multicolumn{4}{c}{Log (Radial Average)}
    &&\multicolumn{4}{c}{Log (Volume Average)}\\
    \cmidrule(r){2-5}\cmidrule(l){7-10}
    Element& I & II & III & IV && I & II & III & IV \\
    \midrule
    Hydrogen & $-2.738$ & $-0.001$ &  \nodata &  \nodata  && $-1.610$ &
    $-0.011$ &  \nodata  &  \nodata  \\ 
    Helium   & $-1.661$ & $-0.009$ &  \nodata &  \nodata  && $-0.567$ &
    $-0.137$ &  \nodata  &  \nodata  \\ 
    Nitrogen & $-3.045$ & $-0.836$ & $-0.069$ & $-3.605$  && $-1.785$ &
    $-0.270$ & $-0.351$  & $-4.288$  \\ 
    Oxygen   & $-2.822$ & $-0.452$ & $-0.191$ &  \nodata  && $-1.584$ &
    $-0.150$ & $-0.574$  &  \nodata  \\ 
    Neon     & $-2.842$ & $-0.169$ & $-0.494$ &  \nodata  && $-1.815$ &
    $-0.058$ & $-0.960$  &  \nodata  \\ 
    Sulfur   & $-5.322$ & $-1.276$ & $-0.042$ & $-1.420$  && $-4.247$ &
    $-0.597$ & $-0.132$  & $-2.069$  \\ 
    Chlorine & $-4.716$ & $-1.093$ & $-0.041$ & $-2.037$  && $-3.644$ &
    $-0.477$ & $-0.177$  & $-2.689$  \\ 
    Argon    & $-3.585$ & $-1.382$ & $-0.023$ & $-1.996$  && $-2.283$ &
    $-0.490$ & $-0.175$  & $-2.657$  \\ 
    \bottomrule
    \tabnotetext{a}{Luridiana \& Peimbert 2001.}
  \end{tabular}
\end{table*}

\paragraph{Helium.}

In NGC~5461 a non-negligible fraction of the total helium is in neutral form.
According to the Luridiana \& Peimbert (2001) model, such fraction is bracketed by the radial
and the volume averages:
\beq
0.02 < \frac{\mathrm{He^0}}{\mathrm{He}} < 0.27.
\eeq
To estimate the ${\mathrm{He}^0}/{\mathrm{He}}$ amount we will take a weighted average
of the radial and volume average ionization structure presented in
Table~\ref{tab:chem_ab}. The relative weights are 0.84 for the radial value
and 0.16 for the volume average, and were obtained by adjusting the observed
S$^+$/S$^{++}$ ratio to the ratio predicted by the model; this provides a good
approximation since the ionization potential of He~{\sc i} is roughly equal to
that of S~{\sc ii}.  {From} these relative weights we obtain that
${\mathrm{He}^+}/{\mathrm{He}}=0.929$, and combining this value with the mean value of
$y^+$ (see Table~\ref{tab:helium}) we find: \beq
\frac{\mathrm{He}}{\mathrm{H}}={y^+}/0.929 = 0.1009, \eeq or \beq Y=\frac{4\times
{\mathrm{He}}/{\mathrm{H}}}{1+4\times {\mathrm{He}}/{\mathrm{H}}+2\times16\times
{\mathrm{O}}/{\mathrm{H}}}=0.285, \eeq where the assumption has been made that oxygen
represents one half of all the mass in heavy elements, and 0.08 dex were added
to the gaseous value for ${\mathrm{O}}/{\mathrm{H}}$ derived below to take into account
the fraction of oxygen trapped in dust grains.

\paragraph{Oxygen.}\label{sec:oxy5461}

The oxygen abundance was calculated by adding the observed ratios \noIInh{}
and \noIIInh:
\beq
\frac{\mathrm{O}}{\mathrm{H}}=\frac{\mathrm{O}^+ + \mathrm{O}^{++}}{\mathrm{H^+}},
\eeq

${\mathrm{O}^0}$ is not included in the sum since it is expected that 
the fraction of neutral oxygen is very similar to that of neutral hydrogen
due to the charge-exchange reaction $\mathrm{O^+ \!+ H^0 \rightarrow O^0 \!+ H^+}$ 
(see Osterbrock 1989,  and Table~\ref{tab:ioniz_av} of this paper).

For the case of O$^{+}$ and O$^{++}$, the ionic ratios amount to 0.354
and 0.644 (radial averages), or 0.708 and 0.267 (volume averages),
while the corresponding observed values are $\mathrm{O^+/O^{tot}|_{obs}}
=0.352$ and $\mathrm{O^{++}/O^{tot}|_{obs}} =0.612$.  The fact that the
observed values are in much better agreement with the radial than with
the volume averages, reflects the fact that the slits used for the
observations sample only a small part of the region.  For this reason,
we will consider in the following only the radius-weighted ionization
correction factors, and we will drop for simplicity the superscript
$R$, i.e., $\mathit{icf}(\mathrm{X}^i)^\mathrm{LP01}\equiv
\mathit{icf}(\mathrm{X}^i)^{\mathrm{LP01}, R}$.

The neutral oxygen is underpredicted by the model, no matter which
kind of average is considered (but particularly in the case of
the radial average). This fact is a common finding in photoionization
modeling (see, e.g., Stasi\'nska \& Schaerer 1999;
Luridiana \& Peimbert 2001), but from the point of view 
of the computation of the ionization correction factors, it has no consequences
since the O$^0$ fraction is practically equal to the H$^0$ fraction.

\paragraph{Nitrogen.}

Only two nitrogen ionic species, N$^+$ and N$^{++}$, are expected in
this \ion{H}{2} region, and only N$^+$ is observed.  The model of
NGC~5461 gives for the ionization correction factor of N$^+$: \beq
\mathit{icf}(\mathrm{N}^+)^\mathrm{LP01}=6.85, \eeq where the radial
average has been considered (see the previous section).  It is
interesting to compare this value with the one obtained for a ``hot''
region according to the method developed by Mathis \& Rosa (1991):
\beq \mathit{icf}(\mathrm{N}^+)^\mathrm{MR91}=3.31.  \eeq yielding for
the total nitrogen abundance a value 0.32~dex smaller than the one
computed relying on the ionization structure of the model by Luridiana
\& Peimbert (2001).  This difference is probably related---at least
partially---to the consideration of the radius-averaged
$\mathit{icf}$, which is less precise in the case of elements measured
by means of lines observed with the longer slit.  This hypothesis
seems confirmed when one considers that the Mathis \& Rosa (1991)
result is bracketed by the radial and volume-averaged
$\mathit{icf}$'s.

\paragraph{Neon.}

The Luridiana \& Peimbert (2001) model of NGC~5461 gives:
\beq
\mathit{icf}(\mathrm{Ne}^{++})^\mathrm{LP01}=3.12.
\eeq

This value can be compared with the one predicted by the method of
Mathis \& Rosa (1991): \beq
\mathit{icf}(\mathrm{Ne}^{++})^\mathrm{MR91}=1.64, \eeq which gives
again a total neon abundance about a factor of 2 smaller than the one
based on the Luridiana \& Peimbert (2001) model. In this case, the
difference cannot be explained in geometrical terms, as in the case of
nitrogen and sulfur, since the volume-averaged
$\mathit{icf}(\mathrm{Ne}^{++})$ of the model by Luridiana \&
Peimbert, which amounts to 9.12, even further away from the Mathis \&
Rosa (1991) value.  This difference should be studied further.

\paragraph{Sulfur.}

In the case of sulfur, a correction must be applied to the ionic abundances
of S$^+$ and S$^{++}$ to account for the presence of a small fraction of S$^{3+}$
inside the H$^+$ sphere. The Luridiana \& Peimbert (2001) model gives:
\beq
\mathit{icf}(\mathrm{S}^++\mathrm{S}^{++})^\mathrm{LP01}=1.04,
\eeq
yielding the total sulfur abundance listed in Table~\ref{tab:chem_ab}.
This $\mathit{icf}$ is almost equal to 1, implying that the S/H ratio is robust.
In what follows, we explore the additional information for the ionization structure
provided by the S$^+$/S ratio. 

The method by Mathis \& Rosa (1991) gives: \beq
\mathit{icf}(\mathrm{S}^+)^\mathrm{MR91}=12.52, \eeq while the
photoionization model by Luridiana \& Peimbert (2001) predicts \beq
\mathit{icf}(\mathrm{S}^+)^\mathrm{LP01}=18.88.  \eeq If the sulfur
abundance were calculated from the observed S$^+$ ratio only, applying
the method by Mathis \& Rosa (1991), the value $\log
\mathrm{S/O}=-1.40$ would be obtained.  This value is 0.07 dex higher
than the value derived before, obtained by direct observations of both
S$^+$ and S$^{++}$, combined with the $\mathit{icf}(\mathrm{S}^+
\!+\mathrm{S}^{++})$ by Luridiana \& Peimbert (2001).  On the other
hand, the value obtained by correcting the observed $\mathrm{S^+/H^+}$
by means of the $\mathit{icf}(\mathrm{S}^+)$ of Luridiana \& Peimbert
(2001) would be $\log \mathrm{S/O}=-1.22$, a value 0.18~dex higher
than the one in Table~\ref{tab:chem_ab}.  An equivalent way of looking
at the same discrepancy is to consider the $\mathrm{S^+/S^{++}}$
value, which is 0.100 according to observations, and 0.053 according
to the Luridiana \& Peimbert (2001) model.  This discrepancy can be
qualitatively explained when one considers that the S$^+$ abundance
value is based on observations made with a larger slit than the one
used in the case of, e.g., \odos{} and \neb{} (see Luridiana \&
Peimbert 2001): the radial average is thus a better approximation in
the oxygen than in the sulfur case.  This view is supported by
considering that the observed $\mathrm{S^+/S^{++}}$ ratio is bracketed
by the radial average (${\mathrm{S}^{+}}/{\mathrm{S}^{++}}|_R =
0.053$) and the volume average (${\mathrm{S}^{+}}/{\mathrm{S}^{++}}|_V
= 0.253$).  However, as we have seen before, the $\mathit{icf}$'s
predicted by {\sc Cloudy} are in general discrepant from those
predicted by Mathis \& Rosa (1991), so that also in this case further
investigation is needed to assess the origin of the discrepancy.

\paragraph{Chlorine.}

We adopted the $\mathit{icf}$ of the model by Luridiana \& Peimbert
(2001) to determine the total chlorine abundance, which is: \beq
\mathit{icf}(\mathrm{Cl}^{++})^\mathrm{LP01}=1.10, \eeq yielding a
total chlorine abundance $\log \mathrm{Cl/O}=-3.54$.  On the other
hand, the value predicted by the method of Mathis \& Rosa (1991) is:
\beq \mathit{icf}(\mathrm{Cl}^{++})^\mathrm{MR91}=2.22, \eeq which
yields $\log \mathrm{Cl/O}=-3.24$, a value very close to the solar one
(see Table~\ref{tab:chem_ab}).

\paragraph{Argon.}

According to the Luridiana \& Peimbert (2001) model, Ar$^{++}$
accounts for 95\% of the total argon, giving
$\mathit{icf}(\mathrm{Ar}^{++})^\mathrm{LP01}=1.05$ and implying the total
abundance listed in Table~\ref{tab:chem_ab}, whereas the method by
Mathis \& Rosa (1991) gives
$\mathit{icf}(\mathrm{Ar}^{++})^\mathrm{MR91}=1.87$, which would imply for
the total argon abundance a value 0.25~dex higher than the adopted
value.

\subsection{NGC~5471}

For NGC~5471 we do not have a detailed photoionization model
available, thus we shall rely on a photoionization model of
NGC~2363.\footnote{More precisely, the model used is one of a series
  of simple burst models which constitute the extended-burst,
  composite model for NGC~2363.}  NGC~2363 is a low-metallicity,
high-excitation giant \ion{H}{2} region, with a degree of ionization
for the oxygen similar to that of NGC~5471.  The model assumes
$Z=0.20~Z_{\sun}$ and a $t=3.0$~Myr instantaneous burst, which ionizes
a spherical region, made up of two concentric shells of different
densities; see Luridiana et~al.\@ (1999) for further details.  To
improve the approximation, we did not directly use the
$\mathit{icf}$'s obtained from this model, but rather rescaled them to
the actual observed ionic ratios.  This procedure, which will be
described more explicitly in the following sections, allows us to
overcome possible differences related to the temperature and
ionization structure of the two \ion{H}{2} regions.  The chemical
abundances inferred for NGC~5471 are listed in
Table~\ref{tab:chem_ab}, and will be discussed in detail in the
following.

\paragraph{Helium.}

The amount of neutral helium is probably negligible inside NGC~5471,
as implied by its high ionization degree (its ionization degree is
similar to that of M17 where it is found that the amount of neutral helium is
negligible, see Peimbert, Torres-Peimbert, \& Ruiz 1992; Esteban et~al. 1999).
Therefore, we find for the total helium abundance (see Table~\ref{tab:helium}), 
\beq
\frac{\mathrm{He}}{\mathrm{H}}=\frac{\mathrm{He}^+ \!+ \mathrm{He}^{++}}{\mathrm{H^+}} = 0.0868,
\eeq
or
\beq
Y=\frac{4\times {\mathrm{He}}/{\mathrm{H}}}{1+4\times {\mathrm{He}}/{\mathrm{H}}+2\times16\times 
{\mathrm{O}}/{\mathrm{H}}}=0.257
\eeq
where the assumption has been made that oxygen represents one half of all the mass in heavy elements,
and 0.08~dex were added to the gaseous value for ${\mathrm{O}}/{\mathrm{H}}$ derived below
to take into account the fraction of oxygen trapped in dust grains.

\paragraph{Oxygen.}\label{sec:oxy5471}

The oxygen abundance was calculated, as in the case of NGC~5461, by simply adding the
observed ratios $\mathrm{O^+/H^+}$ and $\mathrm{O^{++}/H^+}$, so no assumptions
on the ionization structure are necessary. 

\paragraph{Nitrogen.}
\newcommand\BS{\kern-2ex} 
We determined the nitrogen abundance using the observed
$\mathrm{N^{+}/O^{+}}$ ratio and the $\mathrm{O^{+}/O}$,
$\mathrm{N^{+}/N}$ ratios predicted by the model by Luridiana et~al.\@
(1999): \beq \frac{\mathrm{N}}{\mathrm{O}}=
\left(\frac{\mathrm{N}^+}{\mathrm{O}^+}\right)_\mathrm{\!\!obs}\BS\times
\left(\frac{\mathrm{O}^+}{\mathrm{O}}\>\frac{\mathrm{N}}{\mathrm{N}^+}\right)_\mathrm{\!LPL99}.
\eeq Basically, this relation takes advantage of the similarity in the
ionization structure of the two regions, rescaling to the observed
$\mathrm{O^{+}/O}$ for a better approximation.

This equation yields for the total nitrogen abundance relative to
oxygen the value $\log \mathrm{N/O}=-1.23$~dex.  It is interesting to
compare this value to the one obtained by means of the Mathis \& Rosa
(1991) method.  The criterion of Mathis \& Rosa places NGC~5471 into
the category of ``hot'' \ion{H}{2} regions, so that: \beq
\mathit{icf}(\mathrm{N}^+)^\mathrm{MR91}=5.67, \eeq which, combined with
the observed value of N$^+$ gives $\log {\mathrm{N}}/{\mathrm{O}}=
-1.34$.  On the other hand, with the usual assumption
$\mathrm{N^+/O^+={N}/{O}}$ one would obtain $\log \mathrm{N/O} =
-1.31$~dex.  The difference between these two values and the one
adopted by us is probably related to the non-negligible fraction of
N$^{+3}$ in the nebula, which is implicitly neglected by the empirical
equation.

\paragraph{Neon.}
In the case of {Ne}$^{++}$, the scaling equation is: \beq
\frac{\mathrm{Ne}}{\mathrm{O}}=
\left(\frac{\mathrm{Ne}^{++}}{\mathrm{O}^{++}}\right)_\mathrm{\!\!obs}\BS\times
\left(\frac{\mathrm{O}^{++}}{\mathrm{O}} \>
  \frac{\mathrm{Ne}}{\mathrm{Ne}^{++}} \right)_\mathrm{\!LPL99}, \eeq
since the Ne$^{++}$ coexists with O$^{++}$. We obtain for the total
neon abundance $\log \mathrm{Ne/O}= -0.73$~dex.  This result is
intermediate between the prediction of the Mathis \& Rosa (1991)
method, giving: \beq
\mathit{icf}(\mathrm{Ne}^{++})^\mathrm{MR91}=1.05, \eeq and $\log
\mathrm{Ne/O}= -0.76$, and the usual empirical relation
$\mathrm{Ne/O={Ne}^{++}/{O}^{++}}$, yielding $\log \mathrm{Ne/O}=
-0.70$.

\paragraph{Sulfur.}

The total sulfur abundance can be obtained from the following
relations: \beq \frac{\mathrm{S}}{\mathrm{O}}\approx
\left(\frac{\mathrm{S}^{++}}{\mathrm{O}^{+}}\right)_\mathrm{\!\!obs}\BS\times
\left(\frac{\mathrm{O}^{+}}{\mathrm{O}} \>
  \frac{\mathrm{S}}{\mathrm{S}^{++}}\right)_\mathrm{\!LPL99}, \eeq and
\beq \frac{\mathrm{S}}{\mathrm{O}}\approx
\left(\frac{\mathrm{S}^{++}}{\mathrm{O}^{++}}\right)_\mathrm{\!\!obs}\BS\times
\left(\frac{\mathrm{O}^{++}}{\mathrm{O}} \>
  \frac{\mathrm{S}}{\mathrm{S}^{++}}\right)_\mathrm{\!LPL99}, \eeq
since S$^{++}$ coexists with O$^+$ and O$^{++}$.  We adopted the value
which is obtained by the weighted average of equations~(26) and~(27),
where the weight is determined by the observed O ionization fraction,
and obtained $\log \mathrm{S/O}=-1.43$.

The Mathis \& Rosa (1991) method gives: \beq
\mathit{icf}(\mathrm{S}^{+})^\mathrm{MR91}=17.68, \eeq implying $\log
\mathrm{S/O}=-1.38$, in good agreement with the value derived with our
model.

\paragraph{Chlorine.}

We determined the chlorine abundance with relations analogous to the
ones used for sulfur: \beq \frac{\mathrm{Cl}}{\mathrm{O}}\approx
\left(\frac{\mathrm{Cl}^{++}}{\mathrm{O}^{+}}\right)_\mathrm{\!\!obs}\BS\times
\left(\frac{\mathrm{O}^{+}}{\mathrm{O}} \>
  \frac{\mathrm{Cl}}{\mathrm{Cl}^{++}}\right)_\mathrm{\!LPL99} \eeq
and \beq \frac{\mathrm{Cl}}{\mathrm{O}}\approx
\left(\frac{\mathrm{Cl}^{++}}{\mathrm{O}^{++}}\right)_\mathrm{\!\!obs}\BS\times
\left(\frac{\mathrm{O}^{++}}{\mathrm{O}} \>
  \frac{\mathrm{Cl}}{\mathrm{Cl}^{++}}\right)_\mathrm{\!LPL99}, \eeq
and using the same weighting procedure used for sulfur obtained $\log
\mathrm{Cl/O}=-3.71$.
The Mathis \& Rosa (1991) method gives: \beq
\mathit{icf}(\mathrm{Cl}^{++})^\mathrm{MR91}=12.35, \eeq yielding
$\log \mathrm{Cl/O} = -2.81$, a much higher value than our adopted
value.

\paragraph{Argon.}

A lower limit to the total argon abundance is given by: $\log
\mathrm{Ar/O} > \log \mathrm{(Ar^{++}\!+Ar^{3+}/{O})_{obs}} = -2.43$.
However, a correction should be made in order to account for the
presence of a small fraction of Ar$^{+}$.  The equation based on the
photoionization model gives:
\begin{equation} 
  \frac{\mathrm{Ar}}{\mathrm{O}}=
  \left(\frac{\mathrm{Ar}^+ \!\!+\!
      \mathrm{Ar}^{++}}{\mathrm{O}^{++}}\right)_\mathrm{\!\!obs}\BS\times
  \left(\frac{\mathrm{O}^{++}}{\mathrm{O}} \>
    \frac{\mathrm{Ar}}{\mathrm{Ar}^+ \!\!+\!
      \mathrm{Ar}^{++}}\right)_\mathrm{\!LPL99},
  \rule[-3.5ex]{0pt}{0pt}  
\end{equation}
yielding $\log \mathrm{Ar/O}=-2.38$.  This number can be compared to
the prediction based on the Mathis \& Rosa (1991) method: \beq
\mathit{icf}({\mathrm{Ar}^{++}})=
\left(\frac{\mathrm{Ar}}{\mathrm{Ar}^{++}}\right)_\mathrm{\!MR91}=7.62,
\eeq implying $\log \mathrm{Ar/O}=-1.67$, which is very large,
probably meaning that the Ar$^{++}$ fraction for this object is
underestimated by the method of Mathis \& Rosa (1991).

\section{DISCUSSION AND CONCLUSIONS}\label{sec:con}

We have described new high-resolution spectroscopic data obtained for
the two giant \ion{H}{2} regions NGC~5461 and NGC~5471 in M101.  We
have \hbox{determined} five independent temperatures and four independent
densities, as well as the abundances for the following elements: H,
He, N, O, Ne, S, Cl, and~Ar.

The densities obtained from the forbidden lines are considerably higher
than the root mean square densities, implying strong density
fluctuations with typical values of 0.01 for the filling factor.

The [\ion{Ar}{iv}] density for NGC~5471 seems to be larger than
$N_\mathrm{e}$(\ion{O}{ii}); since the [\ion{Ar}{iv}] lines originate
close to the main ionizing sources, this finding probably indicates
that the main ionizing stars are still located in a higher than
average density region. A similar result was obtained for NGC~2363 by
P\'erez, Gonz\'alez-Delgado, \& V\'{\i}lchez (2001) and Esteban
et~al.\@ (2002).

A detailed comparison of the observed line intensities with those
predicted by the photoionization model of Luridiana \& Peimbert (2001)
for NGC~5461 was made. We found that the observed [O {\sc i}] line
intensities are higher than predicted by the model, indicating that
the object is ionization bounded. A similar argument can be made based
on the [\ion{S}{ii}] lines.

The $\mathit{icf}(\mathrm{N}^{+})$ for NGC~5461 derived from the
photoionization model of Luridiana \& Peimbert (2001) is about a
factor of two higher than that derived by Mathis \& Rosa (1991) and
also about a factor of two higher than that derived from the empirical
relationship given by $\mathrm{N^+/O^+} = \mathrm{N/O}$.  Further
analysis is needed to understand the origin of this discrepancy.

The $\mathit{icf}(\mathrm{Ne}^{++})$ for NGC~5461 derived from the
photoionization model of Luridiana \& Peimbert (2001) is about a
factor of two higher than that derived by Mathis \& Rosa (1991) and
also about a factor of two higher than that derived from the empirical
relationship given by $\mathrm{Ne}^{++}/\mathrm{O^{++}} =
\mathrm{Ne/O}$ (Peimbert \& Costero 1969).  In this case we think that
the problem lies with the model by Luridiana \& Peimbert (2001) for
the following reasons: (a)~probably the best determination for the
$\mathrm{Ne/O}$ ratio of any \ion{H}{2} region is that derived by
Peimbert et al.\@ (1992) for M17, amounting to $-0.71$~dex, (b)~we
expect the Ne/O abundance to be practically the same for NGC~5461,
NGC~5471, and M17 because, to a very good first approximation, these
elements are formed by massive stars and are ejected into the
interstellar medium by supernovae of type~II, and (c)~the
$\mathit{icf}(\mathrm{Ne}^{++})$ for NGC~5471 and M17 are considerably
smaller than for NGC~5461 indicating that the abundances for the first
two objects are more reliable than for NGC~5461. The large Ne/O value
derived for NGC~5461 is probably an upper limit due to the presence of
the charge-exchange reaction $\mathrm{O}^{++} \!+ \mathrm{H}^0
\rightarrow \mathrm{O}^+ \!+ \mathrm{H}^+$ that allows some O$^+$ to
coexist with Ne$^{++}$; this effect will reduce the
$\mathit{icf}(\mathrm{Ne}^{++})$ predicted by the model (see Peimbert,
Luridiana, \& Torres-Peimbert 1995, and references therein).
 
For NGC~5471, the ionization correction factors were computed based on
a photoionization model for a similar \ion{H}{2} region.  Comparison
of the total chemical abundances obtained to those implied by the
Mathis \& Rosa (1991) method highlighted sensitive discrepancies,
particularly in the case of chlorine and argon. The origin of these
discrepancies requires further study.

\vspace*{3\baselineskip}


\begin{thebibliography}{} 

\bibitem{}
Allende-Prieto, C., Lambert, D. L., \& Asplund, M. 2001,
\apj, 556, L63

\bibitem{}
Benjamin, R. A., Skillman, E. D., \& Smits, D. P. 1999,
\apj, 514, 307

\bibitem{}
Brocklehurst, M. 1971, \mnras, 153, 471 

\bibitem{}
Casta\~neda, H. O., V\'{\i}lchez, J. M., \& Copetti, M. V. F. 1992,
\aap , {260}, 370 

\bibitem{}
Christensen-Dalsgaard, J. 1998, \ssr, {85}, 19

\bibitem{}
Copetti, M. V. F., Mallmann, J. A. H., Schmidt, A. A., \& Casta\~neda, H. O. 2000, 
\aap, 357, 621

\bibitem{}
de Vaucoulers, G. 1978, \apj, 223, 351 

\bibitem{}
Esteban, C., Peimbert, M., Torres-Peimbert, S., \& Escalante, V. 1998,
\mnras, 295, 401

\bibitem{}
Esteban, C., Peimbert, M., Torres-Peimbert, S., \& Garc\'{\i}a-Rojas, J. 1999,
RevMexAA, 35, 65

\bibitem{}
Esteban, C., Peimbert, M., Torres-Peimbert, S., \& Rodr\'{\i}guez, M. 2002,
\apj, submitted

\bibitem{}
Evans, I. N. 1986, \apj, 309, 544


\bibitem{}
Ferland, G. J. 1996, 
{Hazy, a Brief Introduction to Cloudy}, University of Kentucky 
Department of Physics and Astronomy Internal report

\bibitem{}
Filippenko, A. V. 1982, 
PASP, 94, 715

\bibitem{}
Garnett, D. R., Shields, G. A., 
Peimbert, M., Torres-Peimbert, S., Skillman, E. D, Dufour, R. J., 
Terlevich, E., Terlevich, R. J. 1999, 
\apj, {513}, 168

\bibitem{}
Grevesse, N., \& {Sauval}, A. J. 1998,
\ssr, 85, 161

\bibitem{}
Hamuy, M., Walker, A.R., Suntzeff, N. B., Gigoux, P., Heathcote, 
S. R., \& Phillips, M. M. 1992, PASP, 104, 677

\bibitem{}
Holweger, H. 2001,
AIP Conference Series~598, Solar and Galactic Composition, 
ed. R.F. Wimmer-Schweingruber (New York: Springer), 23

\bibitem{}
Keenan, F. P., Hibbert, A., Ojha, P. C., \& Conlon, E. S. 1993, 
Phys. Scr., 48, 129



\bibitem{}
Kingdon J. B., \& Ferland, G. J. 1995, \apj, {450}, 691

\bibitem{}
Levine, S., \& Chakrabarty, D. 1994, Publicaciones Internas del IA-UNAM, 
No. MU94-04

\bibitem{}
Luridiana, V., \& Peimbert, M. 2001, \apj, {553}, 633

\bibitem{}
Luridiana, V., Peimbert, M., 
\& Leitherer, C. 1999, \apj, {527}, 110

\bibitem{}
Luridiana, V., Peimbert, A., \& Peimbert, M. 2002, in preparation

\bibitem{}
Martin, C. L. 1997, \apj, 491, 561

\bibitem{}
Mathis, J. S., \& Rosa, M. R. 1991, \aap, {245}, 625

\bibitem{}
McCall, M. L., Rybski, P. M., \& Shields, G. A. 1985, \apjs, {57}, 1

\bibitem{}
Osterbrock, D. E. 1989, Astrophysics of Gaseous Nebulae and Active
Galactic Nuclei  (Mill Valley: University Science Books)



\bibitem{}
Peimbert, A., Peimbert, M., \& Luridiana, V. 2002,
\apj, 565, 668

\bibitem{}
Peimbert, M. 1993, RevMexAA, 27, 9

\bibitem{}
Peimbert, M., \& Costero, R. 1969,
Bol. Obs. Tonantzintla y Tacubaya, {5}, 3

\bibitem{}
Peimbert, M., Luridiana, V.,
\& Torres-Peimbert, S., 1995, RevMexAA, 31, 147

\bibitem{}
Peimbert, M., Peimbert, A., \& Ruiz, M. T. 2000, \apj, 541, 688

\bibitem{}
Peimbert, M., Torres-Peimbert, S., 
\& Ruiz, M. T. 1992, RevMexAA, 24, 155

\bibitem{}
  P\'erez, E., Gonz\'alez-Delgado, R. M., \& V\'{\i}lchez, J. 2001,
  Astrophysics and Space Science Supplement, 277, 83

\bibitem{}
 Rayo, J. F., Peimbert, M., \& Torres-Peimbert, S. 1982, \apj, {255}, 1 

\bibitem{}
Ramsbottom, C. A., Bell, K. L., \& Stafford, R. P. 1996,\linebreak\adjustfinalcols 
Atomic Data and Nuclear Data Tables, 63, 57

\bibitem{}
Robbins, R. R. 1968, \apj, 151, 511


\bibitem{}
 Rosa, M. R., \& Benvenuti, P. 1994, \aap, {1}, 1994

\bibitem{}
Sandage, A., \& Tammann, G. A. 1976, \apj, 210, 7

\bibitem{}
Schaerer, D.,~\& 
Vacca, W.~D.\ 1998, \apj, 497, 618

\bibitem{}
Schuster, W. J. 1982, RevMexAA, 1, 129

\bibitem{}
Shields, G. A. 1986, \pasp, 98, 1072

\bibitem{}
 Skillman, E. D., \& Israel, F. P. 1988, \aap, {203}, 226 

\bibitem{}
 Smith, L.~F.\ 1991, IAU 
Symp.~143, Wolf-Rayet Stars and Interrelations with Other Massive Stars in 
Galaxies, eds. R. Hayes \& D. Milne (Dordrecht: Kluwer), 601

\bibitem{}
Smits, D. P. 1996, \mnras, {278}, 683

\bibitem{}
Stasi\'nska, G., \& Schaerer, D. 1999, A\&A, 351, 72

\bibitem{}
 Storey, P. J., \& Hummer, D. G. 1995, \mnras, {272}, 41

\bibitem{}
Torres-Peimbert, S., Peimbert, M., \& Fierro, J. 1989, \apj, {345}, 186 

\bibitem{}
Vacca, W.~D.\ 1994, \apj, 421, 140

\bibitem{}
Verner, D. A., Verner, E. M., \& Ferland, G. J.\ 1996, Atomic Data and Nuclear Data Tables, 64, 1


\end{thebibliography}
\end{document}